\documentstyle[osa,twocolumn,prl,aps,graphicx]{revtex}  

\begin{document}                
\draft


\wideabs{

\title{Microwave conductivity due to impurity scattering in a d-wave superconductor}

\author{ A.J.~Berlinsky$^1$, D.A.~Bonn$^2$, R.~Harris$^2$ and C. Kallin$^1$}

\address{$^1$Brockhouse Institute for Materials Research and
Department of Physics and Astronomy,\\
McMaster University, Hamilton, Ontario L8S 4M1, Canada }

\address{$^2$Department of Physics and Astronomy, University of British Columbia,6224 Agriculture Rd., Vancouver, B.C. V6T 1Z1, Canada}

\date{\today}

\maketitle

\begin{abstract}                
The self-consistent t-matrix approximation for impurity 
scattering in unconventional superconductors is used to interpret 
recent measurements of the temperature and frequency dependence of 
the microwave conductivity of YBCO crystals below 20K. In this theory, 
the conductivity is expressed in terms of a frequency dependent 
single particle self-energy, determined by the impurity scattering phase 
shift which is small for weak (Born) scattering and approaches $\pi/2$ for unitary scattering.  Inverting this process, microwave conductivity data are used to extract an effective single-particle self-energy and obtain insight into the nature of the operative scattering processes.  It is found that the effective self-energy is well approximated by a constant plus a linear term in frequency with a small positive slope for thermal quasiparticle energies below 20K.  Possible physical origins of this form of self-energy are discussed.
\end{abstract}

}


\narrowtext

Microwave surface resistance measurements on high quality YBCO crystals provided some of the early evidence for unconventional superconductivity in the high $T_c$ cuprates.  The observation of a broad peak in the conductivity versus temperature, well below $T_c$\cite{bonn1}, was interpreted in terms of a rapid drop in the scattering rate due to the disappearance of the inelastic scattering which completely dominates the transport near $T_c$.  At the lowest temperatures, the conductivity is determined by scattering due to static disorder. In the context of a generalized two-fluid model, the low frequency conductivity is proportional to the product of a scattering time, $\tau$, and a density of thermally excited charge carriers, $n_n(T)$.  This conductivity was observed to decrease roughly linearly with $T$.  Based on this observation, it was conjectured\cite{bonn2} that $\tau$ saturated at some large low temperature value, implying that $n_n(T)$ varies roughly linearly with $T$.  Subsequently it was observed\cite{hardy} that the penetration depth (and hence the superfluid density $n_s(T)=n_s(0)-n_n(T)$) varies linearly at low $T$, so that the low $T$ behavior of $\sigma(T)$ and $n_n(T)$ are consistent with a constant scattering rate.

However, the conjectured temperature independent low temperature scattering time is difficult to understand from a theoretical perspective.\cite{HPS} When vertex corrections are neglected, the transport scattering rate, $1/\tau$, is essentially equal to twice the one-electron self-energy.  Within the context of the self-consistent t-matrix approximation for quasiparticles with line nodes, the one-electron self-energy is not expected to be constant at low energies,\cite{HPS,HWSEP} and hence one would expect the low temperature transport scattering rate to be temperature dependent. Quite recently, the UBC Group has reported measurements of the temperature dependence of the microwave conductivity of a high purity YBCO crystal at five frequencies in the range of 1 to 75 GHz.\cite{hosseini} These measurements both confirm and extend their earlier results.  Not only are the observed temperature dependences consistent with an almost frequency and temperature independent scattering rate in the impurity-dominated regime below 20K, but the frequency dependence of the conductivity at fixed temperature is consistent with this same scattering rate.  

The confirmation of the simple picture of an energy independent scattering rate is problematic and raises questions about our understanding of the nature of excitations from the high $T_c$ superconducting ground state. This is particularly so because the crystals under study are so clean that one might expect theories based on perturbation theory to apply.  Nevertheless we will show that the weak energy dependence of the scattering rate obtained from the microwave conductivity is inconsistent with standard simple pictures of scattering of d-wave quasiparticles from point impurities, and we will discuss various possible physical interpretations of this energy dependence. We also suggest that the introduction of small quantities of Zn impurities could be used to test the predictions of standard ``dirty d-wave'' theory.

We begin by comparing the microwave conductivity data to the standard model written in terms of the energy dependent single particle lifetime.  Following the work of Hirschfeld {\it et al.}\cite{HWSEP} the conductivity may be written as:
\begin{eqnarray}
\label{cond1}
\sigma_{xx}&(&\Omega,T)={ne^2\over m^*\Delta}\nonumber\\
&\times&\int_{-\infty}^\infty d\omega\left({\tanh\bigl({\beta\omega\over 2}\bigr)-\tanh\bigl({\beta(\omega-\Omega)\over 2}\bigr)\over 2\Omega}\right)
{F(\omega,\Omega)},
\end{eqnarray}
where $\Omega$ is the microwave frequency. The prefactor, $ne^2/ m^*\Delta$, is related to the temperature dependence of the penetration depth, $1/(\mu_0\lambda_{xx}^2(T))=n_s(T)e^2/m^*$. The integral over $\Omega$ of $\sigma_{xx}(\Omega,T)-\sigma_{xx}(\Omega,0)$ is equal to $1/(\mu_0\lambda_{xx}^2(0))-1/(\mu_0\lambda_{xx}^2(T))$. Thus the prefactor can be determined from the slope of the inverse penetration depth at low $T$.  The result is $ne^2/ m^*\Delta=-(\ln 2/2\mu_0)\partial (1/\lambda_{xx}^2(T))/\partial T\approx 10^6\Omega^{-1}$m$^{-1}$ for the a-axis conductivity data of Ref.~\cite{hosseini}. 

In order to address the question of whether the data can be described within the self-consistent t-matrix approximation we need an expression for the function $F(\omega,\Omega)$ for a general form of the quasiparticle self energy.  Given this expression, one can, in principle, invert the data to extract the energy-dependence of the scattering rate, using Eq. (\ref{cond1}). We have derived an expression for the function $F(\omega,\Omega)$ which is similar to that of Ref. \cite{HWSEP}.  Working in the ``node approximation'', expanding the quasiparticle dispersion relations around the four d-wave nodes, and neglecting the real part of the quasiparticle self energy, we find 
\begin{eqnarray}
\label{F}
F(\omega,\Omega)&=&{1\over 2\pi}{\rm Re}\Biggl\{{2\omega-\Omega+i[\Gamma(\omega)-\Gamma(\omega-\Omega)]\over \Omega+i[\Gamma(\omega)+\Gamma(\omega-\Omega)]}\nonumber\\
&\quad&\times\left[\log\left({\omega-\Omega-i\Gamma(\omega-\Omega)\over\omega+i\Gamma(\omega)}\right)+i\pi\right]\nonumber\\
&-&{2\omega-\Omega+i[\Gamma(\omega)+\Gamma(\omega-\Omega)]\over \Omega+i[\Gamma(\omega)-\Gamma(\omega-\Omega)]}\nonumber\\
&\quad&\times\log\left({\omega-\Omega+i\Gamma(\omega-\Omega)\over\omega+i\Gamma(\omega)}\right)\Biggr\}
\end{eqnarray}
where $\Gamma(\omega)$ is the imaginary part of the quasiparticle self-energy.
The $\Omega\rightarrow 0$ limit of this expression is 
\begin{equation}
\label{F0}
F(\omega,0)={\omega\over\pi\Gamma(\omega)}\tan^{-1}{\omega\over\Gamma(\omega)} +{1\over\pi}.
\end{equation}
Inserting Eq. (\ref{F0}) into Eq. (\ref{cond1}) in the limit of  $T\rightarrow 0$ gives $\sigma_{xx}\rightarrow ne^2/\pi m^*\Delta$, the universal limit\cite{palee}, provided that the first term in Eq. (\ref{F0}) vanishes when $\omega\rightarrow 0$. Assuming that $\Gamma(\omega)$ and $\Omega$ are both much less than $\omega$ gives Eq.~(\ref{Fsimp}) below with $1/\tau=2\Gamma(\omega)$.

$F(\omega,\Omega)$  is a complicated function of the two frequencies and of the complex quasiparticle self-energies.  Hirschfeld et al.\cite{HPS}  derived the simplified form, 
\begin{equation}
\label{Fsimp}
F(\omega,\Omega)={\rm Im}[|\omega|/(\Omega-i/\tau(\omega))]
\end{equation}
which is a good approximation when the microwave frequency and the scattering rate are both small compared to $T$ and all three energies are small compared to $\Delta$. 

\begin{center}
\begin{figure}
\includegraphics[width=8.6cm]{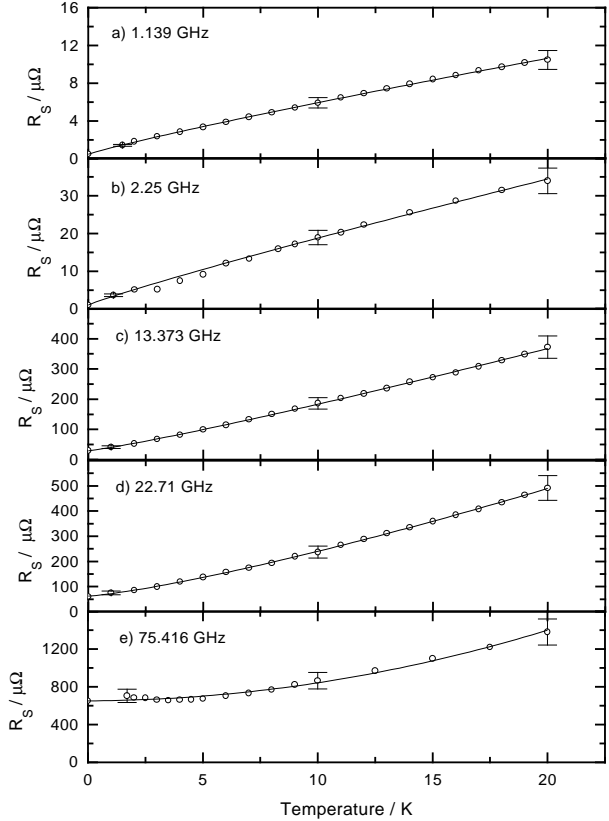}
\caption{Quadratic fits to the surface resistance data of Hosseini {\it et al.} for $T \le$ 20 K.}
\label{data}
\end{figure}
\end{center}

\begin{center}
\begin{figure}
\includegraphics[width=8.6cm]{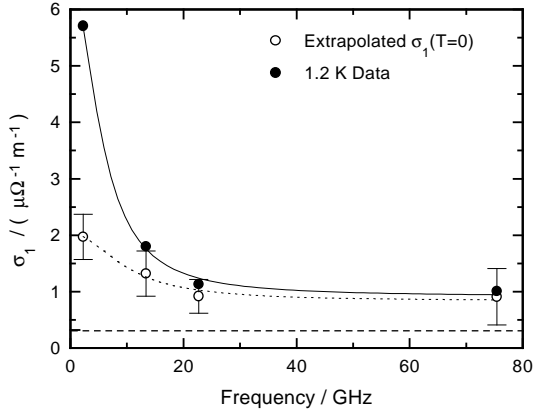}
\caption{The $T=0$ extrapolations of the surface impedance data converted to conductivities.  Shown for comparison are
the 1.2 K data.  The dashed line indicates the predicted universal limit.}
\label{zeroT}
\end{figure}
\end{center}

Surface resistance measurements are converted to conductivities, using the expression
\begin{equation}
R_s(\omega,T)={1\over 2}\mu_0^2\omega^2\lambda^3(\omega,T)\sigma_1(\omega,T)
\end{equation}
The temperature dependence of the penetration depth at 1 GHz is taken from measurements made at that frequency.  For higher frequencies there is a small frequency-dependent contribution to the penetration depth which we obtain self-consistently from our fits to the point-scattering model. The surface resistance data of Hosseini {\it et al.}\cite{hosseini} for $T \le$ 20 K are reproduced in Fig. \ref{data}.  The solid lines in the figure are quadratic fits which allow interpolation and extrapolation to $T=0$.  The $T=0$ extrapolations of these data, converted to conductivities, are shown in Fig. \ref{zeroT}.\cite{comment} All of the extrapolated data are substantially larger than the expected ``universal limit''\cite{palee}, and it has been suggested\cite{durst} that this discrepancy is a real effect due to vertex corrections which enhance the universal conductivity at $T=0$.  In the Born limit, it is enhanced  by a factor of $(\tau_{tr}/\tau)^2$ where $\tau_{tr}$ is the transport lifetime.  The physical origin of these corrections is the fact that, at low $T$, scattering within a node does not change the electric current whereas internode scattering does.  The vertex corrections take the momentum dependence of the scattering potential into account by correctly weighting scattering among the nodes. 

Experimentally it appears that, in addition to a large correction at low $T$, there is also a frequency dependence to the $T=0$ limit, with the conductivity exhibiting a low frequency peak and a slow roll-off at higher frequencies. Theoretically, it is unclear what to expect for the frequency dependence of the conductivity at $T=0$.  At present there are no calculations available for the frequency dependence of the vertex corrections. If vertex corrections are neglected, then one expects the $T=0$ conductivity to roll off at a frequency which depends on the type of scattering.  For unitary scattering, the roll-off is at a rather high frequency, some fraction of $T_c$.\cite{HPS2} For a frequency-independent scattering rate, $1/\tau$, we find that the $T=0$ conductivity falls to half its zero frequency value at about $2.6/ \tau$. Thus, this frequency dependence seems to be sensitive to the detailed nature of the scattering mechanism.
 
\begin{center}
\begin{figure}
\includegraphics[width=8.6cm]{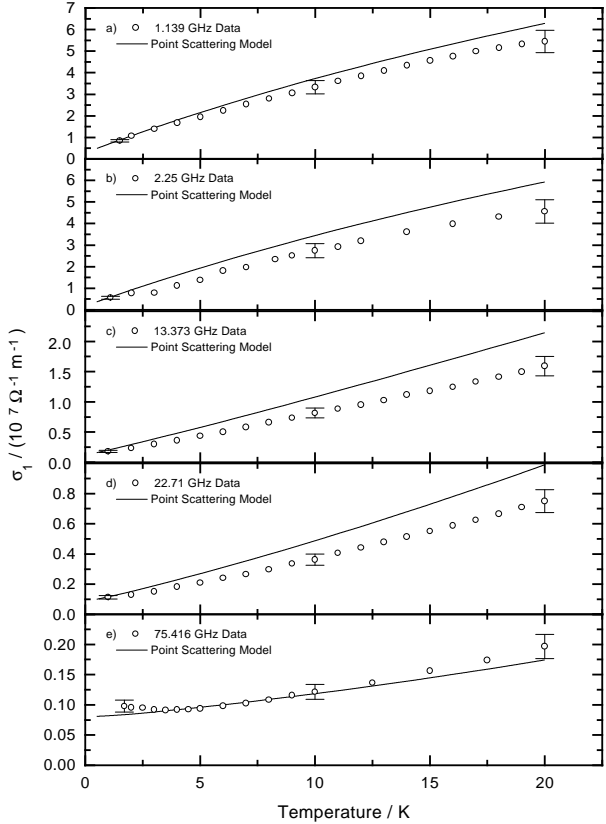}
\caption{The best global fit to the temperature-dependent part of the conductivity below 20 K.}
\label{condata}
\end{figure}
\end{center}

The data for the temperature dependence of the microwave conductivity are shown in  Fig. \ref{condata}.  The data are approximately linear in $T$ at all 5 frequencies, although there is a small downward curvature in the lowest frequency data and a comparable upward curvature at the highest frequency in addition to the clear non-zero intercepts.  
The result of Eq. (\ref{cond1}) for a frequency independent scattering rate, using the simplified form, Eq.~(\ref{Fsimp}) for $F(\omega,\Omega)$ given above, is
\begin{equation}
\label{cond2}
\sigma_{xx}(\Omega,T)={ne^2\over m^*}{\tau\over 1+\Omega^2\tau^2}{(2\ln 2)T\over\Delta}
\end{equation}
For the low frequency data, $\sigma_{xx}$ at 20 K is about 50$ne^2/m^*\Delta$ which implies that $1/\tau$ is  about 0.5 K. 

This semi-quantitative fit to a temperature-\break independent scattering rate is unsatisfying in light of our theoretical picture of the scattering of d-wave quasiparticles.  If the scattering is weak, as might be expected for crystals exhibiting quasiparticle mean free paths of microns,  then one would expect a scattering rate linear in $\omega$ (Born scattering).  Indeed the factor of $|\omega|$ in $F(\omega,\Omega)$ arises from the same source that would lead to a linear scattering rate, namely the linear quasiparticle density of states.   Alternatively, if the scattering is due to a very small density of unitary scatterers, then theory predicts a scattering rate proportional to $1/\omega$.  Of course, there is always the somewhat unnatural possibility that some intermediate type of scattering might  account for the data.  The self-consistent equation for the quasiparticle self-energy depends on the density of impurities and on their phase shift $\delta$, where $\delta <<  1$ corresponds to Born scattering and $\delta \approx \pi/2$ to unitary scattering.    We have examined the possibility that an intermediate value of the phase shift could give rise to an effectively energy independent scattering rate.  What we find is that, regardless of the value of $\delta$, a quasiparticle self-energy with a magnitude on the order of $10^{-3}\Delta$ will have an energy dependence over the relevant range of energies, $0 < \omega < \Delta/10$, which is inconsistent with a constant scattering rate.

Since the data show noticeable curvature and hence do not follow Eq.~(\ref{cond2}) perfectly, we have fit the temperature dependent part of the microwave conductivity using a more general, frequency dependent form of $\Gamma(\omega)$. We model $\Gamma(\omega)$ as a linear combination of weak (Born) and strong (unitary) scattering, and we also allow for the possibility of a frequency independent (Drude) component. Thus $\Gamma(\omega)$ is written as the sum of a linear term, $\Gamma_B\omega/\Delta$, plus a term of the form $\Gamma_u\Delta/\sqrt{\omega^2+\Gamma_u\Delta}$ which mimics the effect of unitary scattering, plus a constant,$\Gamma_D$.  We found that it made little difference in the fits whether Eq.~(\ref{F}) or (\ref{Fsimp}) was used, and so we worked with the simpler expression, Eq.~(\ref{Fsimp}), to fit the temperature-dependent part of the conductivity (with $\sigma(\Omega, T=0)$ subtracted off).  The best global fits, in which data at all frequencies and temperatures were fit to a single set of parameters, are shown by the solid lines in Fig.~(\ref{condata}). The fits provide a reasonable model of the evolution of the curvature of $\sigma$ vs. $T$ as a function microwave frequency.  The values of the parameters which gave the best global fit were: $\Gamma_u=0$, $\Gamma_D=0.33$ K, and $\Gamma_B=0.69$ K.

\begin{center}
\begin{figure}
\includegraphics[width=8.6cm]{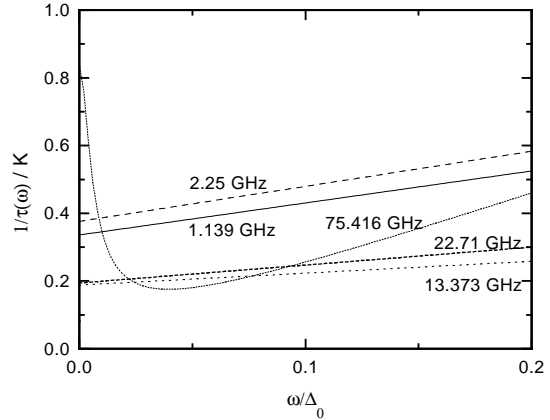}
\caption{The frequency dependence of the one-particle self-energy, as obtained from fits to individual data sets.  The data is best described by a $constant$ + $linear$ self-energy.}
\label{Sigmas}
\end{figure}
\end{center}

In addition we performed individual fits to the temperature dependence of the conductivity at each frequency with our model.  The results for the best-fit parameters are shown in Table (\ref{fits}).
Only the 75GHz data were compatible with a non-zero value of the unitary scattering parameter, $\Gamma_u$ and this fit also required a large value of the linear parameter, $\Gamma_B$.  The frequency dependence of the one-particle self-energy, corresponding to each of these fits, is shown in Fig.~(\ref{Sigmas}). We conclude that the one-particle self-energy consists of a constant term plus a small but clearly positive slope. The  most notable features of this conclusion are the apparent absence of a resonant peak centered at zero frequency that one would expect from unitary scatterers, and the increase of the self-energy with increasing frequency.

\begin{table}
\begin{tabular}{|c|c|c|c|} \hline
Frequency /GHz &  $ \Gamma_{u}$ /K   &  $ \Gamma_{D}$ /K  &  $ \Gamma_{B}$ /K    \\ \hline \hline
1.139 & 0 & 0.34 & 1.12  \\ \hline
2.25 & 0 & 0.38 & 1.22  \\ \hline
13.373 & 0 & 0.19 & 0.42  \\ \hline
22.71 & 0 & 0.19 & 0.63  \\ \hline
75.416 & $2.97 \times 10^{-3}$ & 0 & 2.61 \\ \hline
global & 0 & 0.33 & 0.69  \\ \hline
\end{tabular}
\smallskip
\caption{Best fit parameters to individual data sets using the point scattering model.}
\label{fits}
\end{table}

It is relevant to review the common rationale for expecting unitary scattering to dominate the transport properties in YBCO.\cite{HG}  Early measurements on both crystals and films generally exhibited a relatively flat temperature dependence at low $T$, which reinforced the then commonly held belief that the gap had an s-wave symmetry, although the temperature dependence was more quadratic than exponential.\cite{AGR,bonn2}  The observation of a linear temperature dependence of the penetration depth by Hardy and co-workers\cite{hardy} in very clean samples of optimally doped YBCO, which provided strong evidence for a d-wave gap, also demanded some consistent explanation of the earlier results. Hirschfeld and Goldenfeld\cite{HG} pointed out that, if the extra impurities in earlier samples were unitary scatterers, then a d-wave gap was consistent with the approximately quadratic temperature dependence observed at low $T$ in these materials. By comparison, the amount of Born scattering required to give the observed range of quadratic temperature dependence would also be expected to lower $T_c$ drastically.  They also argued that unitary scattering accounted for the observed linear dependence of the low $T$ thermal conductivity in dirty samples.

Although the arguments given above account for the behavior of dirty samples, i.e. samples with a significant temperature range over which the penetration depth is quadratic in $T$, it is not at all clear that they should apply to the samples used in these experiments, which are extremely pure and which exhibit a linear temperature dependence of the penetration depth down to the lowest temperature studied. As we have seen, the transport scattering rate in these materials is less than 0.5 K at low $T$, and the contribution obtained from extrapolating the linear term in the scattering rate, from the fits above, to $T_c$ would be less than an additional $0.5 K$. These rates are completely negligible compared to the inelastic scattering rate at $T_c$ which is roughly 100 K.  It seems inappropriate to attribute this small amount of scattering to a single mechanism in these samples where the low $T$ quasiparticle mean free path is several microns. More likely the scattering is due to a combination of mechanisms, including rare strong-scattering by point defects, some more ubiquitous weak long-range Born scattering, due to oxygen disorder and weakly scattering lattice defects, and scattering by extended defects such as remnant twin boundaries. Even different kinds of point impurities may have distinctly different kinds of scattering properties.  This is clear from studies of deliberate impurity doping, where a single type of impurity dominates the transport. Work in this area has shown that although Zn appears to behave as a unitary scatterer, other impurities such as Ni and Ca seem to be much weaker scatterers.\cite{czech}.  Whatever the mechanism or combination of mechanisms in the high purity samples, it appears to result in an almost frequency independent quasiparticle lifetime in the temperature range which has thusfar been probed.

Hettler and Hirschfeld (HH) have recently proposed\cite{HH1,HH2} that the apparently frequency independent part of the self-energy is the result of a resonance that arises away from $\omega=0$ due to suppression of the energy gap around the impurity. This finite frequency resonance is superimposed on the zero frequency resonance normally associated with unitary scatterers in d-wave superconductors.  They associate the gap relaxation resonance with the frequency-independent self-energy that we infer from the data.  We agree that gap relaxation must occur around impurities, and that it will affect the quasiparticle self-energy.  It is less clear that it will give rise to a second resonance.\cite{Shnirman}  In any case the effect of the zero frequency resonance which remains in HH's calculation would be to suppress the conductivity at low $T$ in a way that is somewhat inconsistent with our data.  Specifically, the lowest temperature ($T < 5$ K) data at 1 GHz have the opposite curvature to that of HH's fit.

In conclusion, we find that, in the simple picture in which the microwave conductivity is determined by the imaginary part of the quasiparticle self-energy, the self-energy for low-energy quasiparticles acts roughly like a constant plus a small linear term with positive slope, and this behavior is inconsistent with existing theories of a single kind of point-impurity scattering.  Below 20 K, the constant is less than 0.5 K.  We attribute this behavior to the cumulative effect of scattering by a variety of dilute and/or weak scatterers.  The way to test this hypothesis is to introduce a single type of controlled disorder, such as Ni or Zn impurities. These added impurities will cause $R_s$ to decrease and to exhibit a frequency and temperature dependence characteristic of the added impurity. The fact that the starting materials are extremely pure, means that the doping required to do this can be very small, so that the experiments will probe the effects of well-separated dopant atoms. Such experiments may also provide further insight into the nature of vertex corrections which depend on the momentum dependence of the impurity potential.

The authors gratefully acknowledge many useful discussions with W. N. Hardy, A. Hosseini, P. Dosanjh, S. Kamal, A. Durst, P.A. Lee and P. J. Hirschfeld.  This work was initiated during a visit to the Aspen Center for Physics.  It was supported in part by grants from the Natural Sciences and Engineering Research Council of Canada and by the Superconductivity Program of the Canadian Institute for Advanced Research.




\end{document}